\documentclass[9pt,twocolumn,twoside]{pnas-new}
\templatetype{pnasresearcharticle} % Choose template
\title{Beyond Crease Geometry: Multistability in Origami-Inspired Structures through Local Fold Architectures}

\author[a]{Sagi Senderovich}
\author[a]{Ezra Ben-Abu}
\author[a]{Shai Elbaz}
\author[a]{Nadav Zemah}
\author[a]{Anna Zigelman}
\author[a,2]{Amir D. Gat}

\affil[a]{Faculty of Mechanical Engineering, Technion–Israel Institute of Technology, Haifa 3200003, Israel}

\leadauthor{Senderovich}

\begin{document}

\significancestatement{Conventional origami relies predominantly on crease-network geometry to program multistability, linking broader morphing capabilities to greater geometric complexity. We show that another design dimension lies in the nonlinear mechanics of the folds themselves. As a result, local fold parameters govern global deformation and stability, while redistributing bistability within the same crease network shifts global behavior between compliant, spatially distributed phase transformation and discrete switching among hierarchically organized stable configurations. This demonstrates that local fold architecture can determine not only which shapes a structure retains, but also how it moves between them. This local-to-global design principle expands the design space for multistable metamaterials, adaptive structures, and soft robotic systems.}

% Please include corresponding author, author contribution and author declaration information
\authordeclaration{The authors declare no conflict of interest.}
\correspondingauthor{\textsuperscript{2}To whom correspondence should be addressed. E-mail: amirgat@technion.ac.il}

% At least three keywords are required at submission. Please provide three to five keywords, separated by the pipe symbol.
\keywords{Multistable origami $|$ origami tubes $|$  deployable structures $|$  reconfigurable metamaterials}

\begin{abstract}
Origami-inspired tubular structures provide a versatile platform for shape morphing, with multistability achieved predominantly through crease-network geometry. Expanding the range of morphing behaviors can therefore require increasingly intricate crease patterns that become more difficult to model and fabricate, ultimately constraining the realizable morphing landscape. Here, we expand the design space of origami-inspired structures beyond geometry by introducing localized instabilities within the crease network, thereby creating compliant multistable structures whose local fold architectures govern global deformation and stability through both constitutive mechanics and geometric constraints. To relate local fold architectures to global multistability, we develop and experimentally validate a modeling framework in which compliant folds are represented as continuous fields that capture spatially varying bistable mechanics. Force- and displacement-controlled design maps demonstrate that global deformation and stability can be programmed through the fold architecture's local parameters. Redistributing bistability within a fixed crease-network topology shifts the global response between compliant, spatially distributed deformation in semi-bistable architectures and discrete transitions within a hierarchically organized space of stable configurations in fully bistable architectures. These results establish a local-to-global design principle for programming both the stable configurations of a structure and the transition pathways connecting them, expanding the design space for multistable metamaterials, adaptive morphing structures, and soft robotic systems.
\end{abstract}
\maketitle
\thispagestyle{firststyle}
\ifthenelse{\boolean{shortarticle}}{\ifthenelse{\boolean{singlecolumn}}{\abscontentformatted}{\abscontent}}{}

\Firstpage
Multistable structures possess multiple stable states separated by energy barriers, allowing them to retain distinct configurations without continuous energy input. Under external actuation, transitions between such states proceed through unstable regions via snap-through events \citep{harne2017harnessing,cao2021bistable,de2024nonlinear}. These features enable robust and programmable shape morphing, a capability often leveraged in adaptive systems \citep{gorissen2020inflatable,tirado2024earthworm,aza2019multistable}, soft robotic actuators \citep{chi2022bistable,rus2015design,polygerinos2017soft,act11110331,li2017fluid}, and metamaterials \citep{breitman2022flows,peretz2022metafluid,hua2024design}.

One prominent route to realizing such programmable multistability is origami-inspired design, in which global deformation emerges from the coupled mechanics and geometry of interconnected facets and creases. The stiffnesses of these facets and creases determine the energetic cost of deformation, while the geometry of the crease network imposes compatibility constraints on the admissible deformation modes \citep{schenk2013geometry,filipov2016origami,li2024origamimeta}. A prominent tubular example is the Kresling pattern, in which a triangulated cylindrical crease network couples axial displacement and twist \citep{guest1994folding,kidambi2020dynamics}.
\Endparasplit

In conventional origami architectures, however, facets and creases typically provide either rigid constraints or locally monostable compliance, such that global multistability emerges primarily from compatibility constraints associated with crease-network geometry \citep{brunck2016elastic, grey2020mechanics, sharma2025programmable, liu2026actuation}. Consequently, expanding the range of morphing behaviors often requires increasingly intricate crease patterns that become more difficult to model and fabricate, ultimately constraining the realizable morphing landscape.

To overcome this limitation, we expand the origami design space by embedding local bistability within compliant crease networks through shell inversion \citep{zhang2017bistable,bende2015geometrically}, thereby introducing fold architectures that govern global deformation and stability through both local nonlinear mechanics and geometric constraints. We refer to the resulting class of tubular structures as Mixed Folded Tubes (MFTs).

This structural concept motivates an examination of how local nonlinear fold mechanics shape the global structural response and how they can be modeled efficiently and robustly for systematic design. Doing so requires a predictive formulation that treats spatially varying fold geometry and nonlinear fold constitutive behavior as design variables governing the global response. Although the finite element method (FEM) and the reduced-order bar-and-hinge (B\&H) formulations have been used to model the multistable mechanics of compliant shell structures \citep{wo2022bending,zhu2020bar,filipov2017bar}, their systematic application to folding architectures with interacting deformation modes and spatially varying nonlinear mechanics can be computationally intensive and numerically challenging. FEM analyses can become computationally expensive and sensitive to convergence near instabilities, whereas B\&H models require detailed facet-and-hinge discretizations and constitutive calibration of the compliant folds. These limitations motivate BiPlan, a specialized formulation that represents fold extension through continuous fields that capture spatially varying bistable mechanics as the dominant deformation mechanism, thereby relating the local mechanics of the fold architecture to global deformation and stability.

Using BiPlan together with experiments, we compare semi-bistable and fully bistable fold architectures that share a common crease-network topology but differ in the distribution of local bistability. This controlled comparison isolates the influence of local constitutive behavior from that of the underlying geometric constraints and reveals how the distribution of local bistability governs global deformation, stability, and the transition pathways connecting stable configurations.

For the semi-bistable architecture, force- and displacement-controlled design maps demonstrate how global deformation and stability can be programmed through local geometric and constitutive parameters. In the fully bistable architecture, interacting bistable folds produce discrete transition pathways within a hierarchically organized space of stable configurations. The two architectures therefore exhibit complementary mechanical behaviors, ranging from compliant, spatially distributed phase transformations in the semi-bistable MFT to discrete transitions between stable configurations in the fully bistable MFT. Together, these results establish local fold architecture as a design strategy for programming both the stable shapes of a structure and the transition pathways connecting them, positioning MFTs as a versatile platform for multistable metamaterials, adaptive morphing structures, and soft robotic systems.

\section*{MFT Fold Architectures: Geometry and Local Mechanics}

\begin{figure*}[t!]
    \centering
    \includegraphics[width=17.8cm]{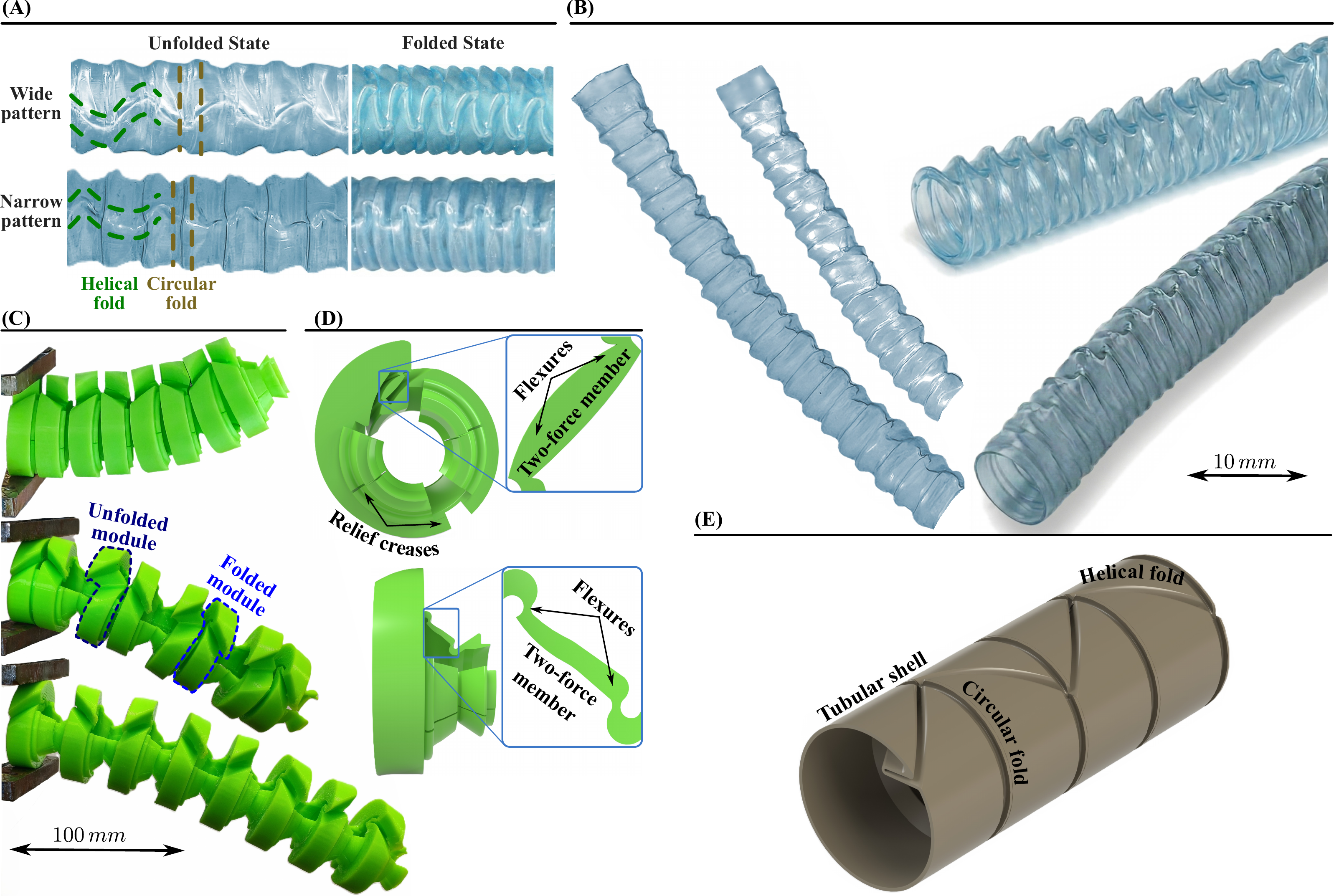}
    \caption{Two Mixed Folded Tube (MFT) variants sharing a helical--circular crease-network topology but differing in their local fold mechanics.
    (A) Wide (top) and narrow (bottom) helical geometries shown in unfolded (left) and folded (right) states. Helical and circular fold trajectories are highlighted by green and brown dashed lines, respectively.
    (B) Corresponding plasticity-induced, semi-bistable MFTs in unfolded and folded configurations.
    (C) Representative deformation states of a flexure-based, 3D-printed fully bistable MFT, with unfolded and folded modules highlighted.
    (D) Front (top) and side (bottom) views of a fully bistable MFT unit, with enlarged views of the corresponding bistable fold mechanisms on the right.
    (E) Schematic of the helical--circular crease network.}
    \label{fig:mft}
\end{figure*}

We consider two MFT variants with a common crease-network topology comprising intersecting helical and circular fold sets embedded in a tubular shell (Fig.~\hyperref[fig:mft]{\ref{fig:mft}.E}). Each variant's fold architecture is defined by the geometry of this crease network together with the constitutive response assigned to each fold set. The crease-network geometry constrains the admissible deformation modes and modulates their geometric coupling. Increasing the circular-fold pitch produces a wider helical pattern with greater axial resistance, whereas decreasing the helical-fold angle produces a narrower pattern with stronger axial–transverse coupling (Fig.~\hyperref[fig:mft]{\ref{fig:mft}.A}). The constitutive response assigned to each fold set then shapes the energetic landscape over this geometry-constrained deformation space.

The first variant employs a semi-bistable fold architecture (Fig.~\hyperref[fig:mft]{\ref{fig:mft}.A--B}) realized through plasticity-induced shell inversion \citep{ben2024directed}. The two fold sets are assigned distinct constitutive responses: the circular folds are bistable, whereas the helical folds remain monostable and approximately linear. The resulting architecture therefore combines local bistability with monotonic axial--transverse coupling.

The second MFT variant employs a fully bistable fold architecture (Fig.~\hyperref[fig:mft]{\ref{fig:mft}.C--D}) realized through 3D-printed folds composed of thickened members connected by thin, eccentric flexural hinges (see \textit{SI Appendix},~S1). This construction localizes buckling along the prescribed fold trajectories and produces bistability in both the circular and helical fold sets.

Thus, the two MFT variants preserve a common helical--circular fold topology while differing in the distribution of local bistability. The semi-bistable architecture confines bistability to the circular folds, whereas the fully bistable architecture distributes it across both coupled fold sets. This controlled variation isolates how the distribution of local bistability governs the global response.

\section*{The BiPlan Modeling Framework}
\begin{figure*}[t!]
    \centering
                \includegraphics[width=17.8cm]{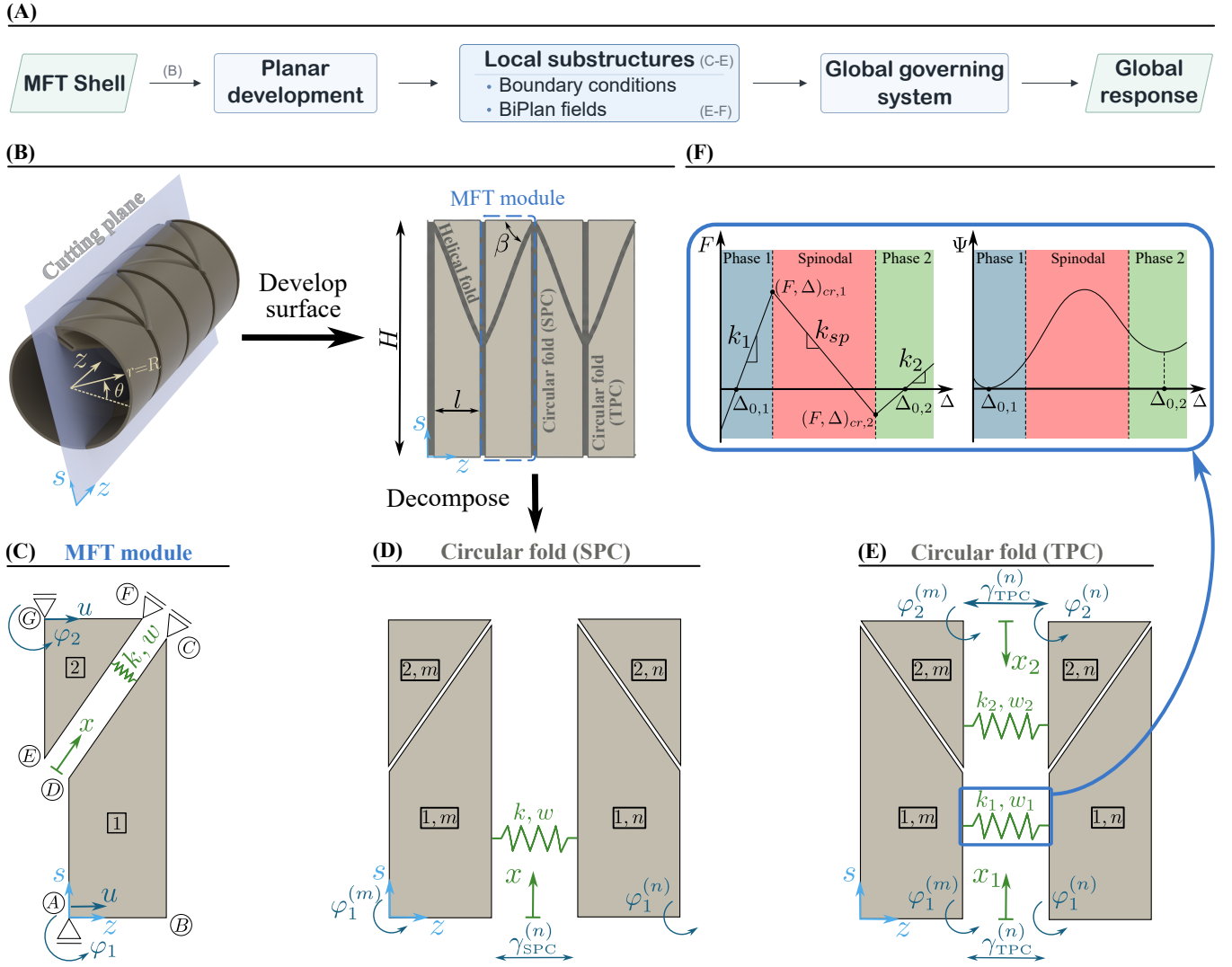}
\caption{Modeling mixed folded tubes (MFTs).
(A) Schematic of the MFT modeling procedure. Indices refer to the corresponding implementation subpanels. (B-E) Implementation of the model procedure to a representative MFT structure.
(B) Planar development of the MFT shell.
(C) Basic MFT module, modeled as two rigid elements connected by a helical compliant crease. Kinematic guides constrain normal motion and enforce compatibility.
(D-E) Local kinematic schemes for SPC and TPC circular-fold connectors between adjacent MFT modules \(m\) and \(n\).
(F) Axial crease compliance represented by continuous bistable stiffness fields, approximated by a trilinear differential force law (left) and the corresponding two-well energy density (right).}
    \label{fig:2}
\end{figure*}

The BiPlan framework, summarized in Fig.~\hyperref[fig:2]{\ref{fig:2}.A} and illustrated for the helical--circular MFT in Figs.~\hyperref[fig:2]{\ref{fig:2}.B--F}, links spatially varying fold geometry and local nonlinear mechanics to global deformation and stability. The MFTs' geometric and mechanical modularity poses two modeling challenges: parameterizing the crease network along the tube and representing bistable constitutive behavior along continuous compliant folds. BiPlan addresses these challenges by mapping the MFT onto a planar domain, decomposing the crease network into characteristic local substructures, and representing compliant folds continuously through bistable planar constitutive fields, termed BiPlan fields.

Unlike lumped hinge or spring models, BiPlan retains the spatial variation of deformation along each continuous compliant fold without introducing independently discretized fold degrees of freedom. The resulting reduced formulation retains fold extension as the leading-order mechanism through which globally dominant axial deformation is accommodated. Accordingly, shell facets are treated as inextensible, fold shear is suppressed through kinematic constraints, and out-of-plane shell bending is neglected. The local field energies are assembled subject to compatibility, guide, and nonpenetration constraints and combined with kinetic energy to determine the dynamic response. Static equilibria are recovered as stationary configurations of the constrained potential, and their local stability is determined from its second variation.

\subsection*{Bistable Planar Constitutive Fields (BiPlan Fields)}\label{sec:2.2}
Within the reduced deformation space defined above, each BiPlan field \(\mathcal{S}\) is compliant along its local opening direction and rigid in the transverse direction. Accordingly, a BiPlan field is characterized by its local elongation \(w=w(x;\boldsymbol q)\) determined by the relative rigid-body motion of the adjacent rigid elements, and the piecewise differential stiffness \(k_{\mathrm{bi}}(w)\) associated with a trilinear constitutive law:
\begin{align}
\label{eq:stiffness}
    \mathcal{S}=\left[w(x;\boldsymbol q),\,k_{\mathrm{bi}}(w)\right], \quad x\in[0,L],
\end{align}
where \(x\) is the local coordinate along a fold axis of length \(L\), and \(\boldsymbol q\) collects the generalized coordinates of the associated substructure (Figs.~\hyperref[fig:2]{\ref{fig:2}.C--E}). The corresponding bistable constitutive relations (Fig.~\hyperref[fig:2]{\ref{fig:2}.F}) exhibit two stable branches separated by an intermediate unstable branch \citep{puglisi2000mechanics}:
\begin{align}
\label{eq:field_constitutive_energy}
F(w) = \int_{0}^{w} k_{\mathrm bi}(\xi)\,d\xi,\quad
\Psi(w) = \int_{0}^{w} F(\xi)\,d\xi,
\end{align}
which, when integrated along the fold axis, yields the strain energy stored in a single BiPlan field:
\begin{align}
\label{eq:field_strain_energy}
\Pi_{\mathcal{S}}=\int_{0}^{L}\Psi\left(w\left(x;\boldsymbol{q}\right)\right)\,dx .
\end{align}
 The spatial variation of \(w(x;q)\) allows different portions of a single fold to occupy different constitutive branches without introducing independently discretized degrees of freedom along the fold, thus enabling continuous phase transformation.
 
 The field's transverse rigidity is imposed by kinematic guides that constrain normal motion along a single boundary (nodes \(C,F\) in Fig.~\hyperref[fig:2]{\ref{fig:2}.C}), while higher-order rotation-induced normal displacements are neglected. The resulting constraints are enforced through guide-axis displacements \(\boldsymbol{g}\) with Lagrange multipliers \(\boldsymbol\lambda\). Additionally, penetration between the connected rigid elements is prevented by constraining the field's boundary elongations to remain nonnegative through unilateral contact conditions (see \textit{SI Appendix},~S2):
\begin{align}
\label{eq:contact}
    \boldsymbol{0}
    \leq
    \boldsymbol{w}_{\partial}
    \perp
    \boldsymbol{\lambda}_{\partial}
    \geq
    \boldsymbol{0},
\end{align}
where \(\boldsymbol w_{\partial}\) and \(\boldsymbol\lambda_{\partial}\) denote the associated boundary elongations and contact Lagrange multipliers, respectively. Thus, the field energy and its kinematic and unilateral constraints are combined in the augmented potential contribution
\begin{align}
\label{eq:aug}
   \Pi_{\mathcal{S},\text{aug}} = \Pi_{\mathcal{S}} + \boldsymbol\lambda^\mathrm T\boldsymbol g - \boldsymbol\lambda_{\partial}^\mathrm T\boldsymbol w_\partial.
\end{align}
\subsection*{Equilibrium Stability}
\label{sec:stability_analysis}

The nonlinear fold-energy landscape may admit multiple global equilibria. At an equilibrium $\boldsymbol q^\ast$, local stability is determined by the second variation of the global constrained potential
\(\Pi_{\mathrm c}\) over perturbations that preserve the active constraints
\(\boldsymbol c(\boldsymbol q)=\boldsymbol 0\). Let
\(\boldsymbol A
=
\left.
\partial_{\boldsymbol q}\boldsymbol c
\right|_{\boldsymbol q^\ast}
\) denote the active-constraint Jacobian, and let \(\boldsymbol N\) span
its null space. The reduced Hessian is therefore
\begin{align}
    \boldsymbol H_r
    =
    \boldsymbol N^{\mathrm T}
    \left.
    \partial_{\boldsymbol q\boldsymbol q}^{2}
    \Pi_{\mathrm c}
    \right|_{\boldsymbol q^\ast}
    \boldsymbol N.
\end{align}

An equilibrium is locally stable when \(\boldsymbol H_r\) is positive definite and loses stability when its smallest eigenvalue reaches zero. The MFT response is classified as monostable, bistable, or multistable according to the number of admissible stable equilibria under the prescribed conditions.

\subsection*{Dynamic Response and Branch Selection}
Multistable fold architectures may induce global stability loss, driving the MFT away from its current equilibrium branch. Although stationary energy conditions identify the admissible equilibria, they do not determine the transient trajectory or the branch dynamically reached under a prescribed loading history \citep{thompson1963basic,shilkrut1994ways}. To resolve this transient evolution and determine the dynamically accessible response branch, BiPlan incorporates the substructure's inertia through the augmented Lagrangian \(\mathcal{L}_{\mathrm{aug}}=T-\Pi_{\mathrm{aug}}\).

The corresponding constrained equations of motion are integrated under the prescribed loading history. Following loss of stability, the resulting transient evolution selects the dynamically accessible deformation path and subsequent response branch (see \textit{SI Appendix},~S3--4).

\subsection*{Application to Helical--Circular MFTs}
We apply the BiPlan framework to the helical--circular crease topology shared by the semi- and fully bistable MFT variants (Fig.~\hyperref[fig:mft]{\ref{fig:mft}}), with detailed geometric and kinematic mappings provided in the \textit{SI Appendix},~ S2--6. The cylindrical shell of
radius \(R\) is developed from \((z,\theta)\) onto the planar domain \((z,s)\), where \(s=R\theta\) (Fig.~\hyperref[fig:2]{\ref{fig:2}.B}). In this domain, the crease
network is parameterized by its circumference \(H=2\pi R\),
circular-fold pitch \(l\), and helical-fold angle
\(\beta\).

The development is decomposed into local MFT modules, each comprising two rigid elements connected by a helical BiPlan field (Fig.~\hyperref[fig:2]{\ref{fig:2}.C}). The motion of module \(n\) is represented by
\begin{align}
\label{eq:2}
\boldsymbol{q}^{(n)}=\left[u_n,\,\varphi_1^{(n)},\, \varphi_2^{(n)}\right]^{\mathrm T},
\end{align}
where \(u_n\) is the common axial translation and \(\varphi_i^{(n)}\) is the in-plane rotation of element $i$. Transverse crease motion is suppressed by kinematic guides, while development compatibility between the corresponding boundaries $AB$ and $GF$ is enforced through the common axial translation $u_n$ and vertical guides. Together, these constraints leave one admissible module degree of freedom (DOF).

Adjacent modules are coupled by circular folds with topology-dependent kinematics (Fig.~\hyperref[fig:2]{\ref{fig:2}.D--E}). A single-pair connector (SPC) comprises a single BiPlan field and forms an open kinematic chain, whereas a two-pair connector (TPC) comprises two parallel fields and forms a closed kinematic loop. The fields' elongations vary with the adjacent modules' rotations, thereby compliantly coupling neighboring modules. The modules’ relative rigid-body displacement determines the boundary elongations:
\begin{align}
    \gamma_{\scriptstyle\mathrm{SPC}}^{(n)}=w(0),\quad
   \gamma_{\scriptstyle\mathrm{TPC}}^{(n)}=w_1(0)=w_2(0),
\end{align}
where the TPC equality enforces development compatibility between its parallel
fields. The MFT is globally constructed by assembling the local generalized coordinates via chain kinematics, enabling the application of global loading and constraints with a global strain energy of
\begin{align}
\label{eq:mft_energy_assembly}
\begin{aligned}
\Pi_{\mathrm{MFT}}
={}&
\sum_{n=1}^{N}
\Pi_{\mathrm{helic}}^{(n)}
\left(\boldsymbol q^{(n)}\right)
\\
&+
\sum_{n=1}^{N}
\Pi_{\mathrm{circ}}^{(n)}
\left(
\boldsymbol q^{(n-1,n)},
\gamma^{(n)}
\right),
\end{aligned}
\end{align}
where \(\boldsymbol q^{(m,n)}\) collects the coordinates
of adjacent modules \(m\) and \(n\), with \(\boldsymbol q^{(0)}\) prescribed by the base boundary conditions. \(\gamma^{(n)}\) denotes the connector-specific coordinate, and helical fold and circular connector contributions are \(\Pi_{\mathrm{helic}}\) and \(\Pi_{\mathrm{circ}}\), respectively. The corresponding inertial and unilateral contact terms are assembled through the shared coordinates and incorporated into the governing system. The two MFT variants share this formulation but differ in their local constitutive assignments: the semi-bistable variant assigns bistability to the circular folds while maintaining linearity in the helical folds through a single-branch differential constitutive law, whereas the fully bistable variant assigns bistability to both fold sets. This common formulation isolates how the spatial distribution of local bistability within a fixed crease topology governs the global dynamic response and equilibrium stability.

\section*{Programmable Deformation and Stability in Semi-Bistable MFTs}
\begin{figure*}
    \centering
    \includegraphics[width=17.8cm]{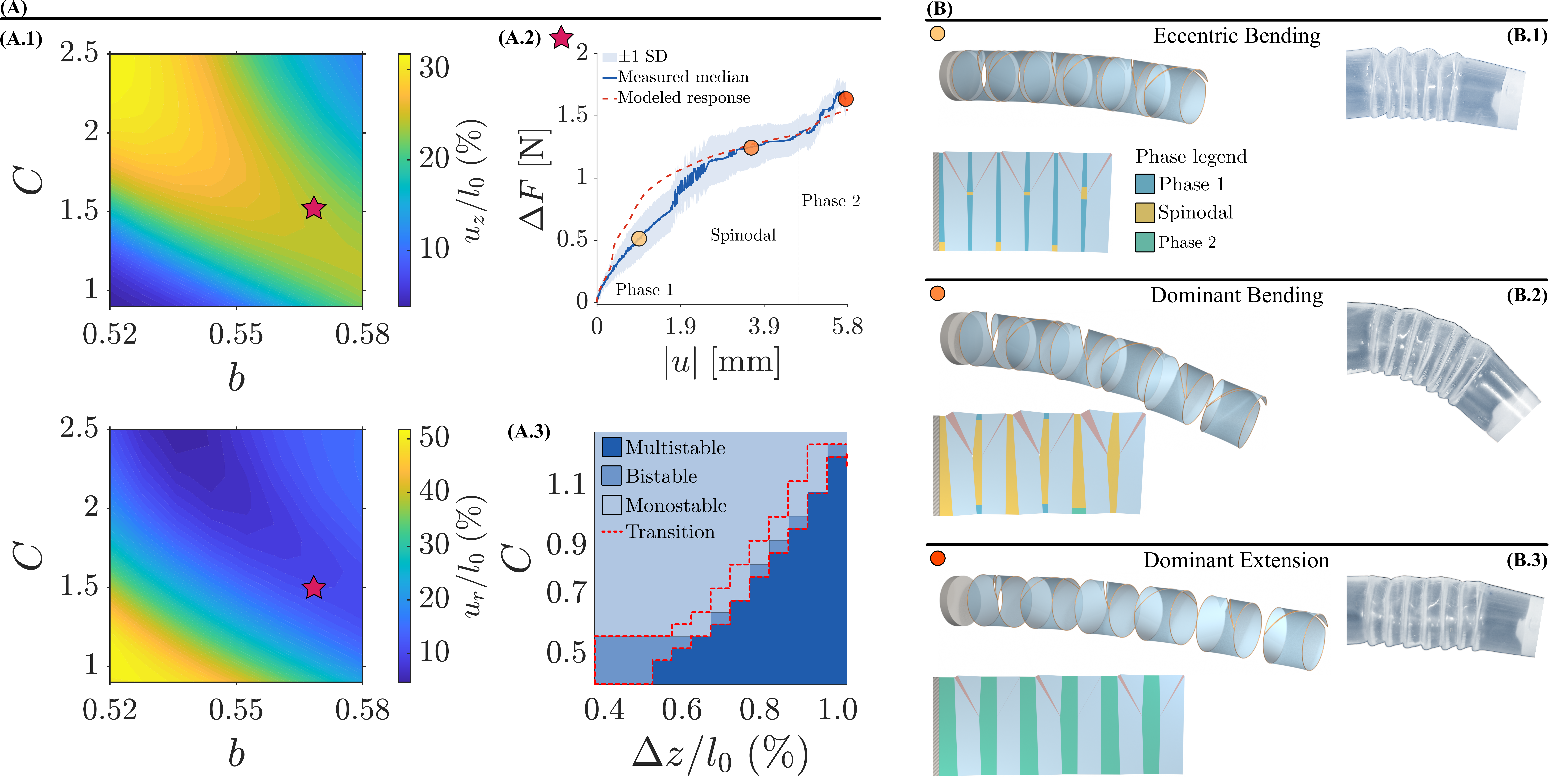}
    \caption{Design-space prediction and experimental validation of a semi-bistable MFT. (A.1) Predicted axial and radial endpoint displacements,
\(u_z\) (top) and \(u_r\) (bottom), normalized by the initial
MFT length \(l_0\), across the \((b,C)\) design plane under a
monotonic axial-force ramp to \(F_f=1.6~\mathrm{N}\). A star marks the experimentally validated design. (A.2) Predicted force--displacement response compared with the measured median and shaded \(\pm 1\,\mathrm{SD}\) band across five semi-bistable MFT specimens. (A.3) Stability map at $b=0.58$ as a function of normalized axial displacement and \(C\), showing mono-, bi-, and multistable regions and the intervening transition region. (B.1--B.3) Representative simulated and experimental configurations in Phase~1, the spinodal regime, and Phase~2, respectively. For each state, the simulated cylindrical reconstruction and planar development are shown on the left, while the corresponding experimental configuration is shown on the right.}
    \label{fig:design-space}
\end{figure*}

Applying the BiPlan formulation, we construct the semi-bistable MFT design space by varying the geometric and constitutive properties of its fold architecture. These properties are parameterized by the fold arc-length and differential stiffness ratios
\begin{align}
\label{eq:basis}
b=\frac{\ell^{\text{helic}}}{\ell^{\text{circ}}},\quad
C=\frac{k^{\text{helic}}}{k_{1}^{\text{circ}}},
\end{align}
respectively, where \(\ell^{\text{helic}}=l/\cos\beta\) and \(\ell^{\text{circ}}=H\). A force-controlled sweep over \(b\) and \(C\) maps the resulting axial and radial endpoint deformations (Fig.~\hyperref[fig:design-space]{\ref{fig:design-space}.A.1}), while a complementary displacement-controlled sweep maps the MFT's stability regimes as functions of the imposed displacement and stiffness ratio \(C\) (Fig.~\hyperref[fig:design-space]{\ref{fig:design-space}.A.3}). The loading conditions and fixed parameter values used in these sweeps are provided in the \textit{SI Appendix},~S9.

As shown in Fig.~\hyperref[fig:design-space]{\ref{fig:design-space}.A.1},
the distinct axial and radial response landscapes show that global MFT
deformation can be biased toward axial extension or radial deflection
through the geometric and constitutive parameters of its fold
architecture. The oblique contours of the axial-displacement map indicate a coupled dependence on \(b\) and \(C\), with maximum extension occurring at low values of both parameters. In contrast, radial displacement decreases from a maximum at low \(b\) and \(C\) toward a minimum at high \(C\) and intermediate \(b\), where radial deformation is most strongly suppressed.

To further examine the predicted MFT response and validate the
BiPlan model, we select a representative design marked by the star
in Fig.~\hyperref[fig:design-space]{\ref{fig:design-space}.A.1} and compare
its force--displacement response with measurements from pneumatically
actuated MFT specimens
(Fig.~\hyperref[fig:design-space]{\ref{fig:design-space}.A.2} and Movie~S1). The constitutive laws assigned to the circular and helical BiPlan fields
were identified independently using measurements from separate Circular
Folded Tube specimens (\textit{SI Appendix},~S10), so the assembled MFT measurements provide an independent validation of the global model response. The prediction agrees well with the experimental measurements, remaining within the \(\pm 1\) standard-deviation band over most of the deformation range, with a slight overprediction of the initial stiffness. BiPlan also captures the overall response sequence, comprising a transition between two stable phases through an intermediate spinodal regime characterized by a loss of global tangent stiffness. To examine the deformation modes associated with each regime, we consider representative states marked along the response curve and shown in Fig.~\hyperref[fig:design-space]{\ref{fig:design-space}.B}.

The simulated and experimental configurations exhibit consistent
phase-dependent deformation modes. During the initial phase, axial loading
extends the linear helical folds together with the bistable circular
folds, while the latter remain predominantly in their first stable state. Coupling through the inclined helical folds causes the circular folds to extend and transition nonuniformly along the circumference, driving global eccentric bending (Fig.~\hyperref[fig:design-space]{\ref{fig:design-space}.B.1}). As a sufficiently large portion of the circular folds enters the spinodal branch, the MFT is driven to global instability, and bending becomes the dominant deformation mode as the circumferential stiffness becomes increasingly nonuniform (Fig.~\hyperref[fig:design-space]{\ref{fig:design-space}.B.2}). Further axial loading completes the transition of the circular folds into their second stable, unfolded state, restoring a globally stable response characterized by dominant extension (Fig.~\hyperref[fig:design-space]{\ref{fig:design-space}.B.3}).

Complementing the force-controlled characterization and model validation, a quasi-static displacement-controlled analysis maps the stability regimes as functions of the imposed displacement and stiffness ratio \(C\) (Fig.~\hyperref[fig:design-space]{\ref{fig:design-space}.A.3}). The map reveals an overall transition from monostability at smaller displacements and higher stiffness ratios to multistability at larger displacements and lower stiffness ratios. These regimes are separated by an intermediate transition region in which the predicted response alternates between bistable and multistable solutions.

Together, the force- and displacement-controlled maps show that the
semi-bistable fold architecture can be tuned to program both endpoint
deformation and the number of stable global equilibria. Because
bistability is confined to the circular folds, coupling to the
monostable helical folds distributes the phase transformation along the
tube, producing a compliant deformation path whose axial and transverse
components can be tuned through local fold geometry and stiffness
contrast.

\section*{Hierarchical State Space and Transition Pathways in Fully Bistable MFTs}
Extending bistability from the circular folds to both coupled fold sets shifts the global response from the distributed phase transformation of the semi-bistable MFT to transitions among discrete stable configurations. Because neighboring modules are kinematically coupled, the state of one fold constrains the deformations compatible with the remainder of the sequence. The resulting stable equilibria form a transition network in which ordered sequences of discrete transitions define pathways between selected initial and target configurations.

\begin{figure*}
    \centering
    \includegraphics[width=17.8cm]{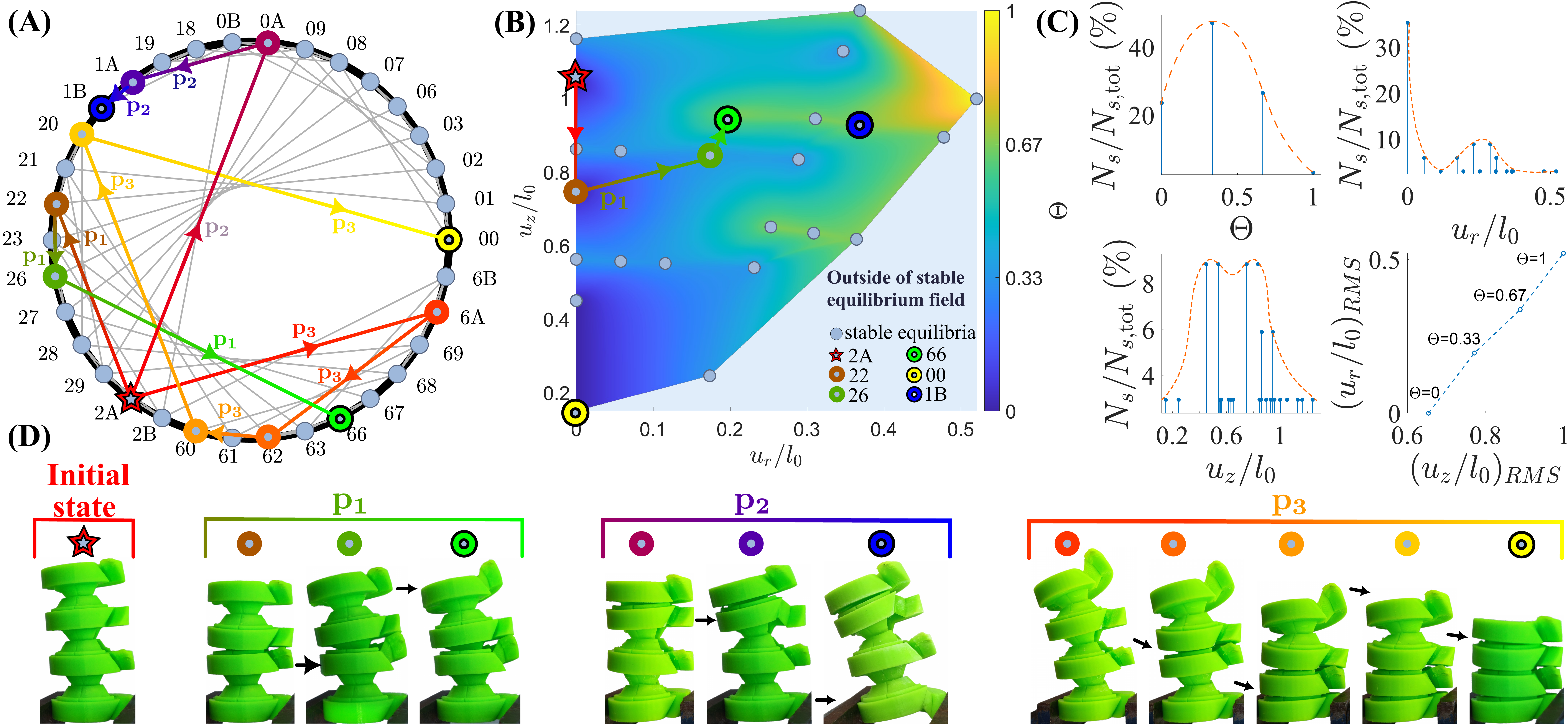}
    \caption{Stable-state landscape of a four-module fully bistable MFT. (A) State-transition diagram constructed from single-fold state
adjacency. Blue nodes denote predicted stable equilibria and are labeled
by two-digit hexadecimal representations of seven-bit fold-state
sequences ordered from the MFT's free end toward its clamped base, with
closed and open folds denoted by \(0\) and \(1\), respectively. A leading
zero is appended before hexadecimal conversion. Gray edges connect
configurations whose fold-state sequences differ by a single bit, and
colored sequences indicate three experimentally realized deformation
pathways. (B) Projection of the stable equilibria into endpoint-configuration space, parametrized by axial and radial endpoint displacements, \(u_z\) and \(u_r\), respectively, normalized by the compressed MFT length \(l_0\), and normalized endpoint slope \(\Theta=\theta/\max|\theta|\). \(p_1\) and target states of \(p_{2,3}\) are marked. (C) state-count distributions over endpoint slope, elongation, and deflection, together with the corresponding branchwise RMS relations. (D) Experimental realizations of the three highlighted deformation pathways. Arrows indicate a fold state change between successive configurations.}
    \label{fig:geo_mft}
\end{figure*}

Using BiPlan, we map the stable equilibria of a four-module fully bistable MFT and construct a transition diagram based on single-fold state adjacency (Fig.~\hyperref[fig:geo_mft]{\ref{fig:geo_mft}.A}; \textit{SI Appendix},~S11). Each node represents a predicted stable equilibrium, while an edge connects two configurations that differ in the state of a single fold. An ordered sequence of connected states defines a deformation pathway between selected initial and target configurations. Projecting the stable equilibria and pathways into endpoint-configuration space, parameterized by elongation \(u_z\), deflection \(u_r\), and slope \(\theta\) (Fig.~\hyperref[fig:geo_mft]{\ref{fig:geo_mft}.B}), reveals a hierarchical distribution of stable states across this space (Fig.~\hyperref[fig:geo_mft]{\ref{fig:geo_mft}.C}). Three representative pathways, highlighted in both the transition diagram and endpoint-configuration space, are reproduced experimentally (Fig.~\hyperref[fig:geo_mft]{\ref{fig:geo_mft}.A--B,D}).

Projecting the stable equilibria into endpoint-configuration space reveals a three-level hierarchy arising from the coupling between the bistable fold states and the MFT chain kinematics (Fig.~\hyperref[fig:geo_mft]{\ref{fig:geo_mft}.B--C}; \textit{SI Appendix},~S12). At the first level, the number of open helical folds sets the accumulated module rotation and therefore determines discrete, equally spaced endpoint-slope branches. Within each slope branch, the circular-fold configuration controls intermodule extension and partitions the states into iso-elongation subbranches. The axial positions of the open helical folds then govern deflection within each subbranch: opening a fold closer to the clamped base rotates a longer downstream portion of the MFT and therefore produces a larger endpoint deflection. Across these hierarchical levels, elongation and deflection also vary systematically with slope, as branchwise RMS averaging shows that they increase approximately linearly with endpoint slope. Thus, helical-fold count, circular-fold configuration, and helical-fold position organize the stable states into a coupled hierarchy in which local fold states successively govern the slope, elongation, and deflection of the global configuration.

The realized pathways demonstrate how this hierarchical state space can be
navigated. Beginning from the common initial state
\(2\mathrm{A}\), pathways \(p_1\) and \(p_2\) terminate at states
\(66\) and \(1\mathrm{B}\), respectively. These target states occupy
the iso-elongation subbranch \(u_z/l_0=0.9\) within the \(\Theta=0.67\) endpoint-slope branch, while exhibiting distinct deflections, consistent with the different axial positions of their open helical folds. Pathway \(p_3\), by contrast, reaches the fully contracted state \(00\), in a different region of the state space. The experiments therefore establish the physical accessibility of selected routes through the state-adjacency graph and show that local fold-switching sequences program both the target configuration and the pathway by which it is reached.

\section*{Concluding Remarks}
In this work, we expanded the design space of origami-inspired structures by introducing localized instabilities within the folds, thereby creating Mixed Folded
Tubes (MFTs), compliant multistable structures whose local fold architectures govern global deformation and stability through both constitutive mechanics and geometric constraints.

BiPlan captures this local-to-global coupling by representing compliant folds through continuous fields that capture spatially varying bistable mechanics, relating local fold geometry and mechanics to global equilibrium configurations, their stability, and the dynamically accessible response following stability loss. For semi-bistable MFTs, the framework predicts the force--displacement response and phase-dependent deformation modes observed in pneumatic experiments. For fully bistable MFTs, BiPlan maps the stable equilibrium states and the transition pathways connecting them, with representative predicted pathways reproduced experimentally.

The design maps demonstrate that global deformation and stability can be programmed through the local geometric and mechanical parameters of the fold architecture, while the controlled comparison between architectures with a common crease-network topology reveals how the distribution of local bistability changes the character of the global response. Restricting bistability to a subset of folds permits compliant deformation through a spatially distributed phase transformation, whereas distributing bistability across coupled folds produces discrete transitions among hierarchically organized stable configurations. Together, these results establish local fold architecture, combining crease-network geometry with local nonlinear mechanics, as a basis for programming both the stable configurations of a structure and the transition pathways connecting them, thereby expanding the realizable morphing landscape for multistable metamaterials, adaptive structures, and soft robotic systems.

\dataavail{All data and code supporting the findings of this study will be
made publicly available on GitHub and archived in Zenodo upon publication.
During peer review, these materials are available from the corresponding
author upon request.}
\bibliography{BIB}
\end{document}

\renewcommand{\thesection}{S\arabic{section}}
\maketitle
\SItext

\section{Fully Bistable MFT -- Design and Fabrication}
\begin{figure}
    \centering
\includegraphics[width=0.5\textwidth]{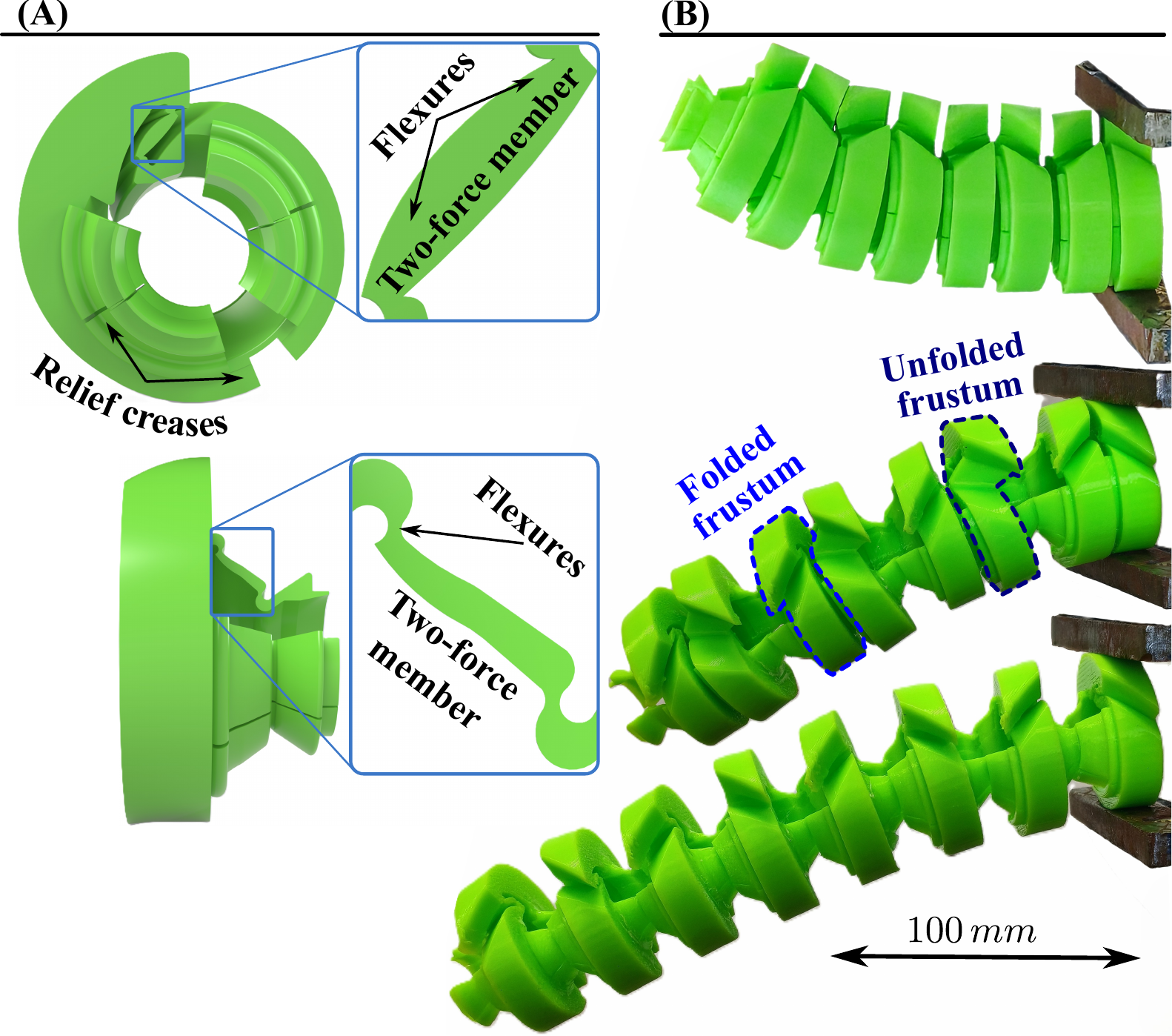}
    \caption{A fully bistable, flexure-based MFT. (A) Front (top) and side (bottom) views of a fully bistable MFT element, with close-ups of the bistable folds (right).
    (B) Representative deformation modes of a flexure-based, 3D-printed MFT, with highlighted frustum states.}
    \label{fig:fdm-mft}
\end{figure}

We fabricate the fully bistable MFT variation by fused deposition modeling (FDM) of individual thermoplastic polyurethane MFT elements (Fig.~\hyperref[fig:fdm-mft]{\ref{fig:fdm-mft}.A}), which are then adhesively bonded into multi-module assemblies (Fig.~\hyperref[fig:fdm-mft]{\ref{fig:fdm-mft}.B}).

 Since FDM prints are prone to interlayer shear failure due to limited adhesion, we use slender hinges to bias deformation toward bending, reducing interlayer shear and improving fatigue resistance. To facilitate shape morphing, kinematic constraints are relaxed by the introduction of relief creases.

\section{Contact Constraints}
The no-penetration condition is imposed on the boundary elongations of each BiPlan. 
For a boundary point \(k\), the scalar gap variable is taken as the boundary elongation 
\(w_{\partial,k}\), so that penetration corresponds to \(w_{\partial,k}<0\). 
Introducing the contact Lagrange multiplier \(\lambda_{\partial,k}\), unilateral contact is written as the complementarity condition
\begin{equation}
    w_{\partial,k}\ge 0,\qquad 
    \lambda_{\partial,k}\ge 0,\qquad
    w_{\partial,k}\lambda_{\partial,k}=0 .
\end{equation}

Thus, either the boundary is open, \(w_{\partial,k}>0\), and the contact force vanishes, or contact is active, \(w_{\partial,k}=0\), and the multiplier prevents penetration. This complementarity condition is regularized using the smoothed Fischer--Burmeister function:
\begin{equation}
\label{eq:sfb}
    \phi_{\mathrm{SFB},k}
    \left(w_{\partial,k},\lambda_{\partial,k}\right)
    =
    \sqrt{
    w_{\partial,k}^{2}
    +
    \lambda_{\partial,k}^{2}
    +
    \varepsilon^{2}
    }
    -
    \left(
    w_{\partial,k}
    +
    \lambda_{\partial,k}
    \right),
\end{equation}

where \(\varepsilon\) is a small smoothing parameter. Collecting all active boundary contact conditions yields
\begin{equation}
    \boldsymbol\phi_{\mathrm{SFB}}
    \left(
    \boldsymbol w_{\partial},
    \boldsymbol\lambda_{\partial}
    \right)
    =
    \boldsymbol 0 .
\end{equation}

\section{Euler-Lagrange Governing Equations}
To incorporate inertia, the kinetic energy $T$ of the associated module is incorporated as
\begin{equation}
\begin{gathered}
    T(\boldsymbol q,\dot{\boldsymbol q})
    =
    \frac{1}{2}
    \dot{\boldsymbol q}^{\mathrm T}
    \boldsymbol M(\boldsymbol q)
    \dot{\boldsymbol q},
    \\
\boldsymbol M(\boldsymbol q)
=
\int_{\mathcal M}
\mu
\left(\partial_{\boldsymbol q}\boldsymbol r\right)^{\mathrm T}
\left(\partial_{\boldsymbol q}\boldsymbol r\right)
\,dA,
\end{gathered}
\end{equation}
where \(\mathcal M\) denotes the shell surface, $\mu$ is the surface density, and \(\boldsymbol r=\boldsymbol r(\boldsymbol r_0;\boldsymbol q)\) is the position of a material point \(\boldsymbol r_0\). Along with the augmented strain energy, these terms produce the module's augmented Lagrangian
\begin{equation}
    \mathcal{L}_{\mathrm{aug}}
    =
    T-\Pi_{\mathrm{aug}}.
\end{equation}
Thus, we obtain each substructure's Euler-Lagrange governing equations and constraints:
\begin{equation}
\begin{gathered}
\label{eq:governing}
    \frac{d}{dt}\left(\partial_{\dot{\boldsymbol \eta}}\mathcal{L}_{\mathrm{aug}}\right) -  \partial_{\boldsymbol \eta}\mathcal{L}_{\mathrm{aug}}= \partial_{\boldsymbol \eta}W_{\mathrm{ext}},\\ 
     \boldsymbol\phi_{\mathrm{SFB}}(\boldsymbol w_{\partial},\boldsymbol\lambda_{\partial})=\boldsymbol 0,
\end{gathered}
\end{equation}
where \(\boldsymbol \eta^\mathrm T = \left[\boldsymbol q^\mathrm T,\, \boldsymbol \lambda^\mathrm T\right]\), and external work \(W_\text{ext}=W_\text{ext}\left(\mathbf{q}\right)\).

\section{Numerical solver}
\label{sec:numerical}
\begin{figure}
    \centering
    \includegraphics[width=\textwidth]{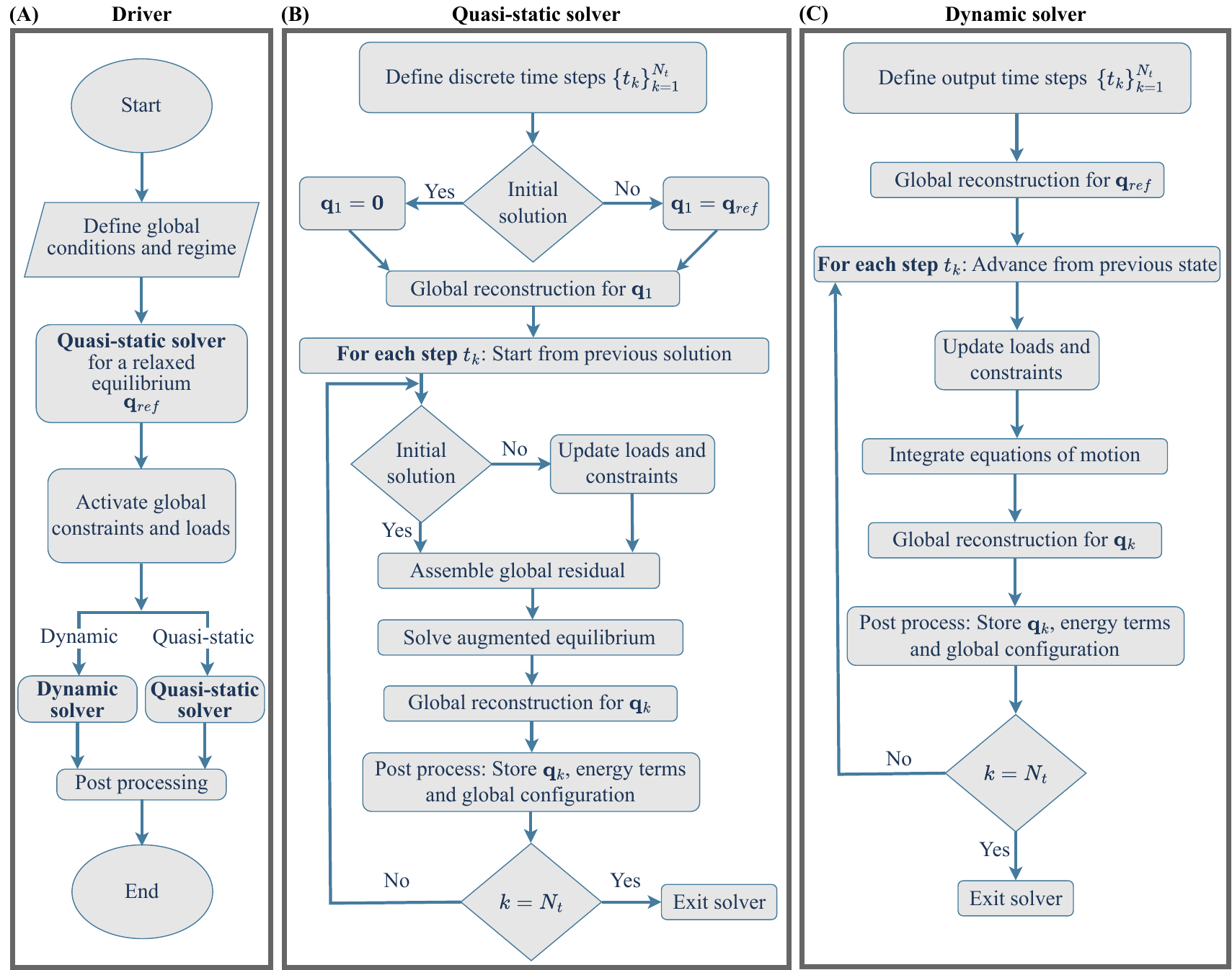}
\caption{Numerical solution scheme for an MFT under prescribed boundary
conditions. (A) The driver defines the global conditions and solution
regime, initializes a relaxed reference configuration, activates the
prescribed loads and constraints, and post-processes the stored
solution. (B) The quasi-static solver incrementally updates the loads
and constraints, assembles the global residual, solves the augmented
equilibrium system, and reconstructs the converged configuration at
each step. (C) The dynamic solver advances the previous state by
updating the loads and constraints and implicitly integrating the
equations of motion, after which the global configuration is reconstructed and stored at each output time.}
    \label{fig:6}
\end{figure}

The quasi-static response of an MFT under prescribed loading and boundary conditions is computed using an incremental Gauss–Newton scheme, whereas its dynamic response under prescribed initial and boundary conditions is obtained using an implicit Newton-based time-integration scheme. Both procedures are summarized in Fig.~\ref{fig:6}.

As shown in Fig.~\ref{fig:6}, the computational procedure consists of
a driver and separate quasi-static and dynamic solvers. The driver
defines the global conditions and solution regime, then invokes the
quasi-static solver to compute a relaxed reference configuration
\(\boldsymbol q_{\mathrm{ref}}\). During this reference solve, the MFT
is subject only to the clamped-base condition. The prescribed
loads and constraints are then activated, and the driver calls the
appropriate solver to compute the constrained response.

Within the quasi-static solver
(Fig.~\hyperref[fig:6]{\ref{fig:6}.B}), the loading interval is
discretized into steps. Each step is initialized
from the previously converged equilibrium, after which the prescribed
loads and constraints are evaluated and the global residual is
assembled from the local governing equations. The resulting augmented
nonlinear algebraic system is solved iteratively for the generalized
coordinates and Lagrange multipliers. Convergence is declared when the
prescribed residual and step-size tolerances are satisfied. The
converged configuration is then reconstructed in the global frame and
stored together with its energy terms before advancing to the next
step. Once the sweep is complete, the solution history is returned to
the driver for post-processing.

Within the dynamic solver
(Fig.~\hyperref[fig:6]{\ref{fig:6}.C}), the response is evaluated at prescribed output times, beginning from the
reference configuration \(\boldsymbol q_{\mathrm{ref}}\). At each step,
the loads and constraints are updated and the equations of motion are
advanced implicitly from the preceding state using a Newton-based
time-integration scheme. The resulting configuration is reconstructed
in the global frame and stored together with its energy terms. This
procedure is repeated over the prescribed time interval, after which
the complete response history is returned to the driver for
post-processing.

\section{Kinematics and Kinetics}
\subsection{Mixed folded tube module}\label{sec:material mapping}
We begin by constructing the kinematic mapping of each rigid element comprising the MFT module within the planar domain. First we define rotation and translation matrices
\begin{equation}
\mathbf{R}(\varphi) = \left[
\begin{array}{cc}
 \cos\varphi & -\sin\varphi  \\
 \sin\varphi  & \cos\varphi \\
\end{array}
\right],\quad \mathbf{t}(\Delta z) = \begin{bmatrix}
    \Delta z\\0
\end{bmatrix},
\end{equation}

 respectively, and an arbitrary material point's initial location vector
 \begin{equation}\mathbf{r}_0^T = \left[z_0,\, s_0\right].\end{equation}
 
In particular, \(\mathbf r_{0,\varphi_i}\) denotes the rotation-center location vector for element \(i\), located at nodes \(A\) and \(G\) for \(i=1\) and \(i=2\), respectively (Fig. 2.C in the main text). The resulting rigid-body mapping is
\begin{equation}
\label{eq:10}
\boldsymbol{\chi}_i(\mathbf q_i,\mathbf r_0)
=\mathbf R(\varphi_i)\bigl(\mathbf r_0-\mathbf r_{0,\varphi_i}\bigr)+\mathbf t(u),
\quad i\in\{1,2\}.
\end{equation}

We apply the established mapping to describe the MFT module's crease elongation \(w(x)\). Since the module's elements are restricted to rigid-body motions, \(w(x)\) varies linearly and is therefore fully determined by its boundary values. Accordingly, we compute the relative displacements of the endpoint node pairs \((D,E)\) and \((C,F)\), project them onto the crease axis, and interpolate the resulting boundary values to obtain \(w(x)\). To this end, we define the relative-displacement operator
\begin{equation}
\boldsymbol{\Delta}_{\partial}=
\begin{bmatrix}
\boldsymbol{\delta}_{DE}^T\\[3pt]
\boldsymbol{\delta}_{CF}^T
\end{bmatrix},
\qquad
\boldsymbol{\delta}_{DE}=\boldsymbol r_D-\boldsymbol r_E,\quad
\boldsymbol{\delta}_{CF}=\boldsymbol r_C-\boldsymbol r_F,
\end{equation}

where the current position of any node \(a\) associated with element \(i\) is
\begin{equation}
\boldsymbol r_a=\boldsymbol\chi_i(\mathbf q_i,\boldsymbol r_{a,0}),
\quad i\in\{1,2\}.
\end{equation}

The corresponding boundary elongations are obtained by projecting the endpoint relative displacements onto the helical crease axis \(\boldsymbol n_{sp}\):
\begin{equation}
\label{eq:bound}
\boldsymbol w_{\partial}=\boldsymbol{\Delta}_{\partial}\boldsymbol n_{sp},
\quad
\boldsymbol n_{sp}^T=[-\sin\beta,\,\cos\beta],
\end{equation}

where the mirrored configuration is obtained by reversing the axis direction, \(\boldsymbol n_{sp}\mapsto-\boldsymbol n_{sp}\). With \(\boldsymbol w_{\partial}^T=:[w_0,\,w_a]\), crease elongation along the axis \(x\in[0,a]\) is given by the linear interpolation
\begin{equation}
\label{eq:17}
w(x)=\left(1-\frac{x}{a}\right)w_0+\frac{x}{a}\,w_a,
\quad x\in[0,\,a],
\end{equation}
where \(a=l/\cos\beta\). Thus, the module's corresponding potential energy is
\begin{equation}
    \Pi_{\text{helic}} = \int_0^a\Psi\left(w(x)\right)dx,
\end{equation}

where \(\Psi\) represents the bistable energy density (Eq. 3 in the main text).

\subsection{Circular crease connector}
The single-pair connector (SPC) and the two-pair connector (TPC) between adjacent MFT modules \(m\) and \(n\) are parametrized by the interface displacements \(\gamma_{\mathrm{SPC}}^{(n)}\) and \(\gamma_{\mathrm{TPC}}^{(n)}\), and by the rotational generalized coordinates \(\varphi_{\ell}^{(m)}\) and \(\varphi_{\ell}^{(n)}\), with \(\ell=1\) for the SPC and \(\ell\in\{1,2\}\) for the TPC (Figs. 2.D-E in the main text). The SPC forms an open kinematic chain and is therefore determined by the configuration of a single neighboring module relative to the connector's local reference configuration. By contrast, the TPC forms a closed kinematic chain, so its elongation is dictated by the relative motion of both modules:
\begin{equation}\label{eq:VC_w}
w(x_\ell)=
\begin{cases}
\gamma_{\mathrm{SPC}}^{(n)}-x_\ell\,\tan\!\bigl(\varphi_{1}^{(n)}\bigr),
& \text{SPC},\ x_\ell\in[0,H],\ \ell=1,\\[2mm]
\gamma_{\mathrm{TPC}}^{(n)}+x_\ell\!\left[\tan\!\bigl(\varphi_{\ell}^{(n)}\bigr)-\tan\!\bigl(\varphi_{\ell}^{(m)}\bigr)\right],
& \text{TPC},\ x_\ell\in[0,H_\ell],\ \ell\in\{1,2\},
\end{cases}
\end{equation}

where \(H_2=l\tan\beta\) and \(H_1=H-H_2\). Thus, the connectors' corresponding potential energy is
\begin{equation}\label{eq:25}
\Pi_{\mathrm{con}}=
\begin{cases}
\displaystyle \int_{0}^{H}\Psi(w(x_\ell))\,dx_\ell, & \text{SPC},\\[3mm]
\displaystyle \sum\limits_{\ell=1}^{2}\int_{0}^{H_\ell}\Psi(w(x_\ell))\,dx_\ell, & \text{TPC}.
\end{cases}
\end{equation}

\section{Constrained displacements}\label{sec:3.1}
We apply the rigid-body mapping functions \eqref{eq:10} to express boundary conditions as algebraic constraints on the mapped nodal positions (Fig. 2.C in the main text):
\begin{equation}
\begin{gathered}
\mathbf{r}_\text{C} = \boldsymbol{\chi}_1(\boldsymbol q_1,\boldsymbol r_{0,C}),\quad 
\boldsymbol{r}_\text{F}=\boldsymbol{\chi}_2(\boldsymbol q_2,\boldsymbol r_{0,F}), \quad \text{where} \;\; \mathbf{r_{0,C}}^T = \mathbf{r_{0,F}}^T = \left[l,\,H\right],
\end{gathered}
\end{equation}

with the corresponding constrained direction \(\boldsymbol{n}_c^T=\left[\cos\beta,\,\sin\beta\right]\) in the standard configuration and by \(-\mathbf n_c\) in the mirrored configuration. We project the displacements of nodes \(C\) and \(F\) onto the constraint direction to obtain the scalar constraint measures
\begin{equation}
\label{eq:11}
g_C=\left(\mathbf r_C-\mathbf r_{0,C}\right)\cdot\boldsymbol n_c,\qquad
g_F=\left(\mathbf r_F-\mathbf r_{0,F}\right)\cdot\boldsymbol n_c,
\end{equation}
which are enforced using Lagrange multipliers, while the compatibility constraints at nodes \(A\) and \(G\) are naturally satisfied by the mapping functions.

\section{Contact Constraints}
The SFB regularization \eqref{eq:sfb} is applied to the boundary elongations of the representative MFT substructures. For the MFT module, no-penetration between rigid elements is enforced at the two endpoints of the helical BiPlan (Fig. 2.C in the main text). Using the boundary elongation vector \(\boldsymbol w_\partial\) defined in \eqref{eq:bound}, the module's contact conditions are
\begin{equation}
\label{eq:fsb_mft}
\boldsymbol\phi_{\mathrm{SFB}}^{\mathrm{mft}}
\left(
\boldsymbol w_\partial,
\boldsymbol\lambda_\partial
\right)
=
\boldsymbol 0.
\end{equation}

For the circular-crease connectors, no-penetration is enforced through the most restrictive boundary elongation over the connector BiPlans:
\begin{equation}
\label{eq:fsb_con}
\phi_{\mathrm{SFB}}^{\mathrm{con}}
\left(
w_\partial,
\lambda_\partial
\right)
=
0,
\quad
w_\partial=
\min_{\ell,\,x_\ell\in\partial\Omega_\ell} w_\ell(x_\ell),
\end{equation}

where \(\ell\) indexes the connector BiPlans and \(\partial\Omega_\ell\) denotes their boundary points.

\section{Global Reconstruction}
We consider a sequence of \(N\) MFT modules, indexed by \(m\in\{1,\dots,N\}\), and their connecting interfaces. Each module contains \(n\) nodes and is parameterized by the generalized-coordinate vectors
\begin{equation}
\mathbf q_i^{(m)}=[u_m,\,\varphi_i^{(m)}]^{\mathrm T},
\quad i\in\{1,2\},
\end{equation}

together with the interface displacements \(\gamma_m\), where interface \(m\) connects modules \(k=m-1\) and \(m\) (Figs. 2.C-E in the main text). For each module \(m\), the local rigid-body mapping \eqref{eq:10} yield the nodal position matrix \(\mathbf X^{(m)}(\mathbf q^{(m)})\in\mathbb R^{2\times n}\) in the \((z,s)\) plane, whose \(j\)th column \(\mathbf X_j^{(m)}\) provides the position of node \(j\). The full sequence is reconstructed by assembling the local module configurations through chain kinematics.

While TPCs introduce a locally closed kinematic loop, SPCs form an open chain in which rotations and deflections propagate to downstream modules. We therefore employ dedicated kinematic operators for each connector type. For adjacent modules \(k\) and \(m\), the SPC relative motion is represented by a rotation \(\theta_m=\varphi_{1}^{(k)}\) about the pivot point \(\mathbf p_m=\mathbf X_1^{(k)}\), yielding the homogeneous rigid-body transforms
\begin{equation}
\mathbf{H}_{m}^{\mathrm{SPC}}=
\begin{bmatrix}
\mathbf{R}(\theta_m) &
\mathbf p_m-\mathbf{R}(\theta_m)\mathbf p_m\\
\mathbf 0 & 1
\end{bmatrix},
\quad
\mathbf{H}_{m}^{\mathrm{TPC}}=\mathbf I_3,
\label{eq:Hu_pivot_min}
\end{equation}

where \(\mathbf H_m\in\left\{\mathbf{H}_{m}^{\mathrm{SPC}},\; \mathbf{H}_{m}^{\mathrm{TPC}}\right\}\) denotes the operator associated with the connector type under consideration. The corresponding translation vector is
\begin{equation}
    \boldsymbol a_m = \Delta_{m}\cdot\begin{bmatrix}
        1,\, c_m
    \end{bmatrix}^T,
\end{equation}

where
\begin{equation}
\Delta_{m} = \gamma_{m}-u_m + l,\quad     c_m = \begin{cases}
        \tan\theta_m,&\quad \text{SPC,}\\[2pt]
        0,&\quad \text{TPC}.
    \end{cases}
\end{equation}

The relative homogeneous transform from module \(m\) to the adjacent upstream module \(k\) is therefore
\begin{equation}
\mathbf{T}_m=
\begin{bmatrix}
\mathbf{I}_{2} & \boldsymbol{a}_m\\
\mathbf{0}^T & 1
\end{bmatrix}
\mathbf{H}_m
\in\mathbb{R}^{3\times 3},
\label{eq:Tu_min}
\end{equation}

with \(m\in\left\{2,\dots,N\right\}\) and $\mathbf{T}_1=\mathbf{I}$. Thus, the absolute placement of module $m$ in the global frame is obtained by the standard serial product 
\begin{equation}
{}^{0}\mathbf{T}_m=\prod_{j=2}^{m}\mathbf{T}_j,
\quad {}^{0}\mathbf{T}_1=\mathbf{I}.
\label{eq:abs_transform_product_min}
\end{equation}

Using the global transform matrix, we lift the local node matrix to homogeneous form:
\begin{equation}
    \tilde{\mathbf X}^{(m)}=
\begin{bmatrix}
\mathbf X^{(m)}\\
\mathbf 1^T
\end{bmatrix}\in\mathbb R^{3\times n},
\end{equation}

and obtain the global homogeneous node coordinates from the single matrix product
\begin{equation}
\bar{\mathbf{X}}^m= {}^{0}\mathbf{T}_m\,\tilde{\mathbf{X}}^{(m)}.
\label{eq:global_module_min}
\end{equation}

Having established the global reconstruction procedure, the external loads and constraints comprising the structure's global boundary conditions can be applied in the global frame via standard virtual work and Lagrange multipliers, respectively.

\section{Semi-Bistable MFT Design Maps}
The force-controlled deformation maps in Fig.~3.A.1 are obtained from
the dynamic response of the semi-bistable MFT under the
monotonic axial-force ramp
\begin{equation}
    F(t)=\dot{F}t,\quad
    \dot{F}=0.8~\mathrm{N\,s^{-1}},\quad
    0\leq t\leq2~\mathrm{s},
\end{equation}
corresponding to a terminal force \(F_f=1.6~\mathrm{N}\). The reported axial and radial endpoint displacements are evaluated at terminal time. The geometric ratio \(b\) and stiffness ratio \(C\) are varied over the ranges shown in Fig.~3.A.1, while all remaining parameters are held at the values listed in Table~\ref{tab:calib}. The geometric ratio $b$ is varied through $\beta$, while $H$ and $l$ remain fixed. The stiffness ratio $C$ is varied through $k_{\text{helic}}$, while the circular-fold constitutive law remains fixed.

The displacement-controlled stability map in Fig.~3.A.3 is computed
quasi-statically at \(b=0.58\). At each prescribed displacement and
stiffness ratio \(C\), distinct equilibria are identified and classified
using the constrained reduced-Hessian stability criterion described in the main text.

\section{Experimental Calibration}
\begin{figure}
    \centering
    \includegraphics[width=\textwidth]{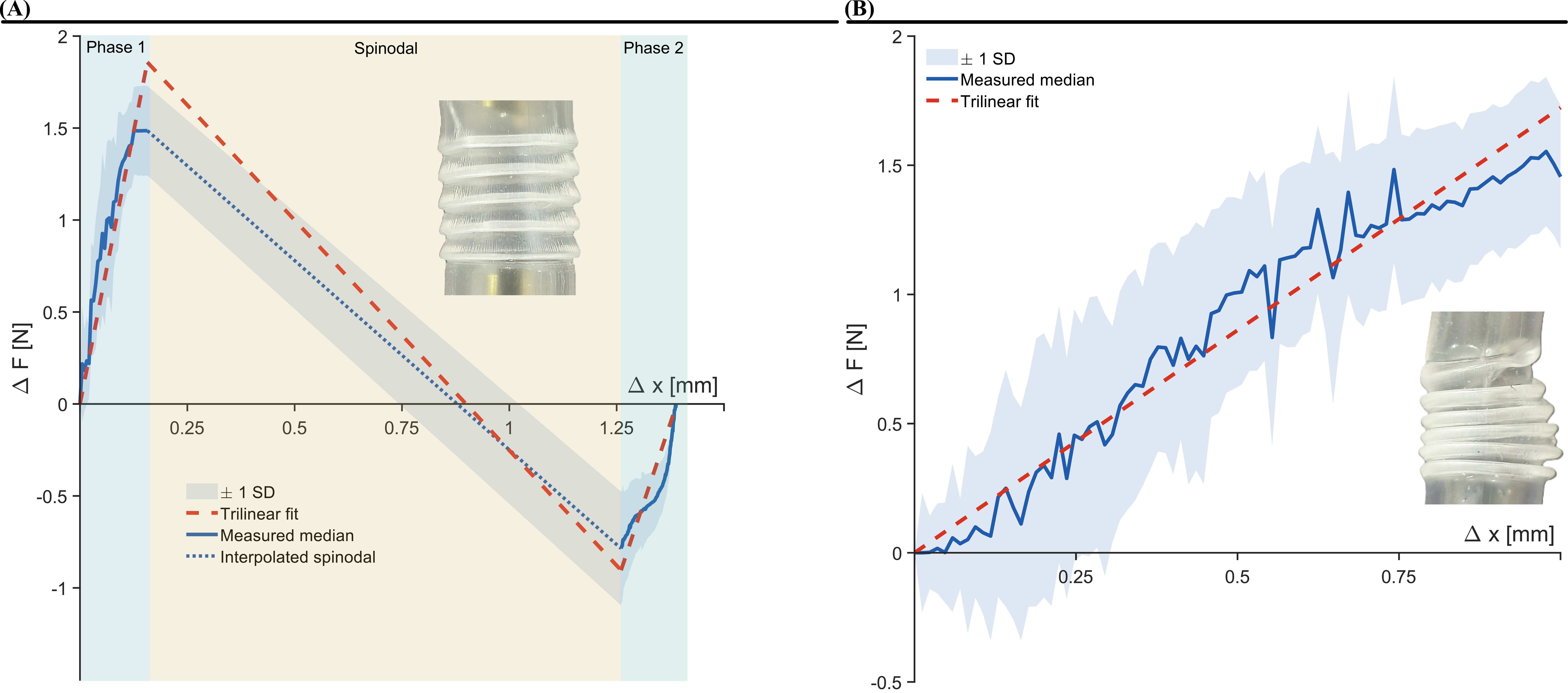}
    \caption{Constitutive responses of Circular Folded Tubes (CFTs) incorporating (A) trilinear-stiffness folds and (B) linear-stiffness folds.}
    \label{fig:calib}
\end{figure}

To validate the proposed model against the experimental response of the semi-bistable MFT, we must first calibrate its stiffness and geometric parameters using measurements from the fabricated specimen. While the geometric parameters can be measured directly, the trilinear and linear stiffness laws governing the circular and helical folds, respectively, contribute simultaneously to the MFT response and therefore cannot be identified independently from measurements of the complete structure. To isolate these contributions, we fabricate Circular Folded Tubes (CFTs) containing either trilinear or linear-stiffness folds and measure their responses under axial loads (Fig.~\ref{fig:calib}).    

The linear CFT is measured under pressure-driven extension. For the trilinear CFT, the first and second phases are characterized under pressure-driven extension and gravity-induced contraction, respectively, up to the stability limit of each phase. The response within the spinodal region is then interpolated between the two measured instability points.

The responses are measured across ten specimens, from which the median response is calculated together with a \(\pm 1\) standard-deviation band. The median responses are then fitted with linear functions using a least-squares approach. These fits are distributed along the folds to define the constitutive laws governing their spatially varying stiffness fields, as specified in Table~\ref{tab:calib}.

\begin{table}
    \centering
    \caption{Geometric and constitutive parameters of the validated
semi-bistable MFT.}
    \label{tab:calib}
    \begin{tabular}{ccc}
        \hline      
        \textbf{Parameter} & \textbf{Symbol} & \textbf{Value} \\
        \hline

        \multicolumn{3}{c}{\textbf{Geometry}} \\
        \hline

        Radius
        & \(R\)
        & \(2.0~\mathrm{mm}\) \\

        Module length
        & \(l\)
        & \(3.6~\mathrm{mm}\) \\

        Helical fold inclination
        & \(\beta\)
        & \(60^\circ\) \\

        Number of modules
        & \(N\)
        & \(6\)\\
        \hline
        \multicolumn{3}{c}{\textbf{Stiffness laws}} \\
        \hline

        Circular-fold stiffness coefficients
        & \(\boldsymbol{k}_{\mathrm{circ}}\)
        & \(\left[0.9,\,-0.2,\,0.6\right]~\mathrm{N/mm^2}\) \\

        Circular-fold transition lengths
        & \(\boldsymbol{l}_{\mathrm{circ}}\)
        & \(\left[0.2,\,1.3\right]~\mathrm{mm}\) \\

        Helical-fold stiffness
        & \(k_{\mathrm{helic}}\)
        & \(1.3~\mathrm{N/mm^2}\) \\

        \hline
    \end{tabular}
\end{table}
\section[Transition Diagram Construction]{Construction of a State Transition Diagram for Fully Bistable MFTs}
The state transition diagram presented in the main text is constructed through a BiPlan stability sweep over all possible fold-state configurations of a four-module fully bistable MFT. The MFT contains seven bistable folds, each of which may occupy either its closed or open stable branch, yielding \(2^7=128\) binary fold-state combinations. For each combination, the local coordinates of the corresponding substructures are initialized toward the prescribed stable branches, and the resulting candidate configuration is relaxed quasi-statically using the procedure described in Sec.~\ref{sec:numerical}. After convergence, each fold is classified as closed or open according to the stable phase occupying the greater fraction of its length. Each converged configuration is then assessed using the constrained reduced-Hessian criterion described in the main text. Equilibria with a positive-definite reduced Hessian are retained as locally stable states, whereas unstable equilibria are discarded.

\begin{figure}
    \centering
    \includegraphics[width=0.5\textwidth]{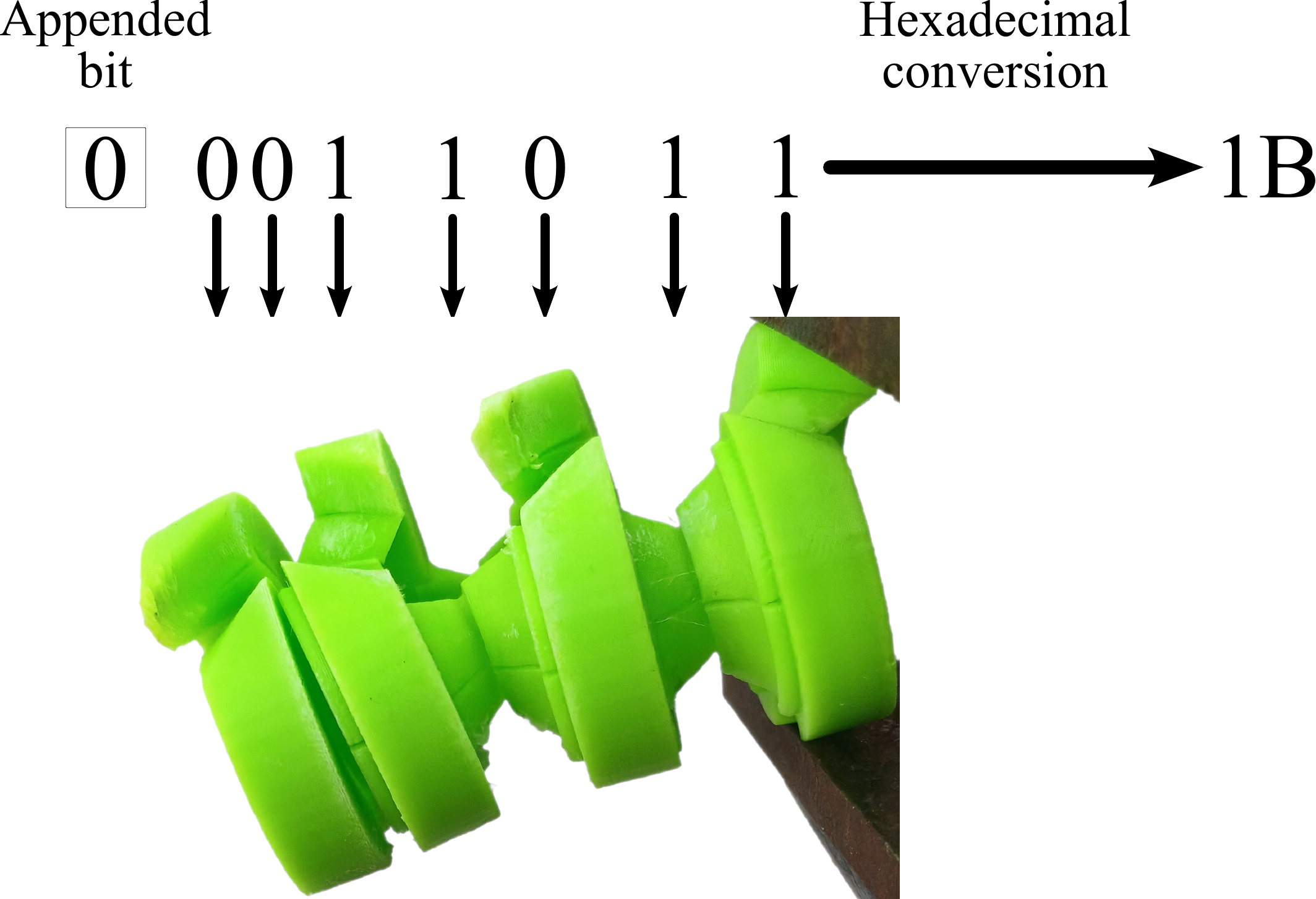}
    \caption{Binary and hexadecimal encoding of a stable fully bistable MFT
configuration. The seven fold states are ordered from the free end toward
the clamped base, with closed and open folds denoted by \(0\) and \(1\),
respectively. A leading zero is appended to form the eight-bit sequence
\(00011011\), corresponding to the hexadecimal label
\(\mathrm{1B}\). The example shows the target configuration of
pathway \(p_2\) in the main text.}
    \label{fig:decode}
\end{figure}

The sweep yields 34 stable equilibria, each of which is represented by a seven-bit binary
sequence ordered from the MFT’s free end toward its clamped base, with closed and open folds denoted by \(0\) and \(1\), respectively. A leading zero is appended to form an eight-bit sequence that can be represented
by a two-digit hexadecimal label. For example, appending a leading zero to the binary sequence \(0011011\) yields \(00011011\), which corresponds to the hexadecimal label \(1\mathrm{B}\) (Fig.~\ref{fig:decode}).
Two retained stable equilibria are connected by an undirected edge when their seven-bit binary sequences have a Hamming distance of one, such that they differ in the state of a single fold. The resulting graph therefore represents single-fold adjacency between stable configurations, and an ordered sequence of connected states defines a transition pathway between selected initial and target configurations.

\section{Configuration-Space Hierarchy of Fully Bistable MFTs}
\begin{figure}
    \centering
    \includegraphics[width=\textwidth]{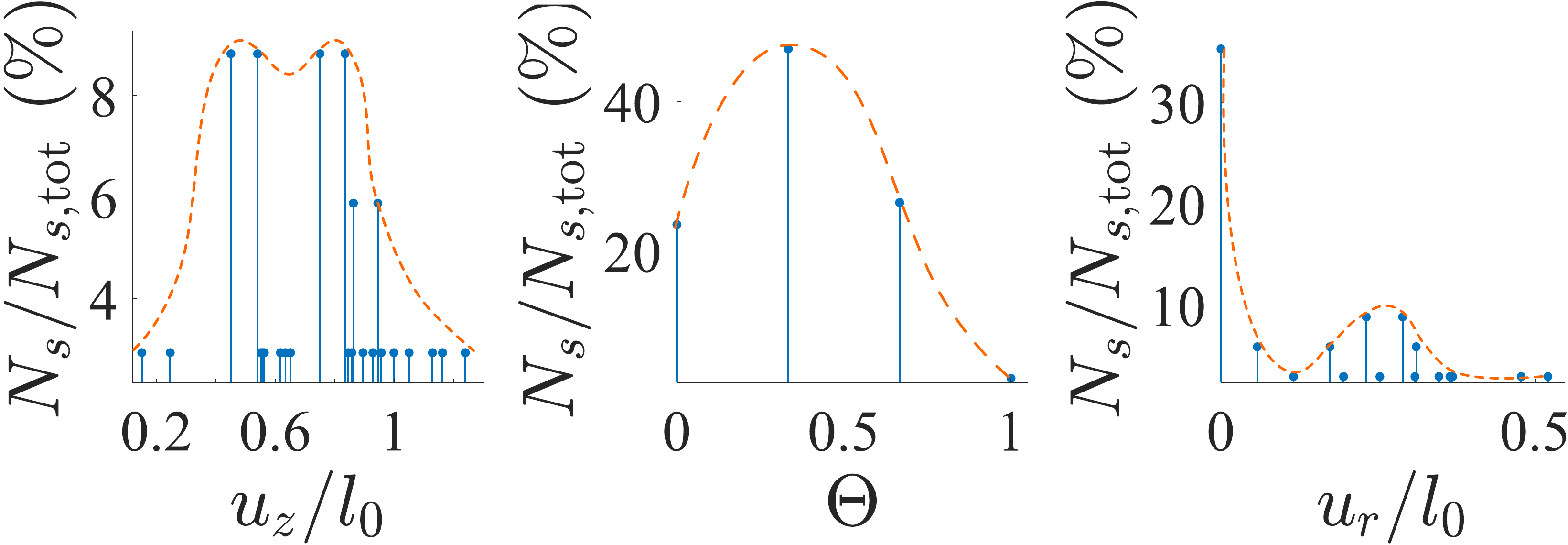}
    \caption{Distribution of stable states within the equilibrium field and the corresponding RMS relations for endpoint deflection, elongation, and slope.}
    \label{fig:dists}
\end{figure}
Figure~\ref{fig:dists} reports equal-weight state-count distributions
over endpoint slope, elongation, and deflection. Because most stable
states contain a single open helical fold, the slope distribution is
unimodal and peaks near \(\Theta=1/3\). The elongation distribution is
bimodal, reflecting the iso-elongation subbranches associated with the
\(\Theta=1/3\) and \(\Theta=2/3\) slope branches. The deflection
distribution exhibits a dominant peak near zero, arising from purely
elongational states and from states in which only the helical fold
nearest the free end is open, such as the first state of pathway
\(p_3\) (Fig.~4.D in the main text). A smaller
secondary peak spans \(u_r/l_0\in[1/8,\,5/8]\), where configurations
with one and two open helical folds produce comparable deflections under
different circular-fold configurations
(Fig.~4.B in the main text).

\FloatBarrier

\movie{Experimental validation of the BiPlan framework for a pneumatically actuated six-module semi-bistable MFT. Top: modeled response in the planar development. Middle: reconstructed cylindrical response. Bottom: experimentally measured response.}